\documentclass[9pt]{article}
\usepackage{authblk} 
\usepackage{amsmath}
\def\frontmatter@preabstractspace{0pt}
\def\frontmatter@postabstractspace{0pt}
\usepackage[version=3]{mhchem} 
\usepackage{braket}
\usepackage{mathtools}
\usepackage{algorithm}
\usepackage{algorithmic}
\usepackage{geometry}
\usepackage{amsmath}
\usepackage{amsthm}
\usepackage{graphicx}
\usepackage{fancyvrb}
\usepackage{forest}
\usepackage{tikz-qtree}
\usepackage[dvipsnames]{xcolor}
\usepackage[english]{babel} 
\usepackage{booktabs}
\usepackage{bm}
\usepackage[style=chem-acs, backend=biber]{biblatex}
\begin{document}

\title{Capturing nuclear quantum effects in high-pressure superconducting hydrides and ice with nuclear--electronic orbital theory}

\author[a]{Logan E. Smith}
\affil[a]{Department of Chemistry, Princeton University, Princeton, New Jersey 08544, United States}
\author[b,c]{Paolo Settembri}
\affil[b]{PSI Center for Scientific Computing, Theory and Data, Paul Scherrer Institute, 5232 Villigen PSI, Switzerland}
\affil[c]{Department of Physical and Chemical Sciences, University of L'Aquila, Via Vetoio, 67100 L'Aquila, Italy}
\author[d]{Alessio Cucciari}
\author[d]{Lilia Boeri}
\affil[d]{Dipartimento di Fisica, Sapienza - Universit\`a di Roma, 00185 Rome, Italy}
\author[c,e]{Gianni Profeta}
\affil[e]{CNR-SPIN L'Aquila, Via Vetoio, 67100 L'Aquila, Italy}
\author[a]{Sharon Hammes-Schiffer}

\maketitle

\begin{abstract}
Nuclear quantum effects are essential for correctly describing hydrogen-rich materials at high pressures. Superconducting hydrides and ice are prime examples of such systems, requiring the inclusion of lattice anharmonicity and nuclear quantum effects to correctly predict and describe the structures and phase transition pressures observed experimentally. Herein, we show that the nuclear–electronic orbital density functional theory (NEO-DFT) method, which treats specified nuclei quantum mechanically on the same level as the electrons, is capable of accurately describing nuclear quantum effects in superconducting hydrides and ice. NEO-DFT predicts the hydrogen-bond symmetrization pressure in H$_3$S and D$_3$S, benchmarking against the more expensive stochastic self-consistent harmonic approximation (SSCHA) method, and predicts the correct symmetric {\it{Fm}}$\Bar{3}m$ structure for LaH$_{10}$ at a wide range of pressures. NEO-DFT also predicts the ice VIII to ice X phase transition pressures for H$_2$O and D$_2$O in agreement with experimental measurements. The accuracy, computational efficiency, and broad applicability of the NEO method opens the door for expanded large-scale studies into these types of systems.
\end{abstract}
\newpage

\section*{Introduction}
\indent As the demand for advanced energy solutions accelerates, solid hydrides have become a research focus, promising breakthroughs from efficient hydrogen storage to superconductors with record-breaking critical temperature ($T_\text{c}$) values \cite{Bellosta_hydrides_appreview_2019, Boeri_JPCM_2021_roadmap}.  Computational methods for crystal structure prediction have facilitated the discovery of novel hydride phases \cite{Parrinello_PRL_2003_MetaDyn1, Godecker_JCP_2004_MHM, Pickard_JPCM_2011_RandomSearch, USPEX_1, USPEX_2, Wang_calypso_2012}, enabling first-principles investigations of phase diagrams for binary, ternary, and even quaternary hydride compounds \cite{Duan_SciRep_2014_SH, Liu_LaH_2017, Zhao_ternary_hydrides_review_2023, DiCataldo_PRB_2021_LaBH8, Ferreira_lunh_NatComm_2023, Conway_hcno_highpressure_2021}.
However, discrepancies between theoretical predictions and experimental results often arise \cite{Boeri_JPCM_2021_roadmap}, complicated by the fact that hydrogen atoms are undetectable by X-ray diffraction, making it difficult to resolve their positions within the crystal lattice unambiguously. These discrepancies are attributed primarily to the inadequate theoretical treatment of lattice anharmonicity and nuclear quantum effects (NQEs) in standard density functional theory (DFT) calculations \cite{Monacelli_JPCM_2021_SSCHA}. The low mass of hydrogen amplifies quantum fluctuations, leading to pronounced delocalization and tunneling that can fundamentally alter the ground state structures of these materials \cite{Mauri_PRL_2015_SH3, Mauri_Nature_2020_LaH, monacelli-2023,Belli_NQEs_2026}. This limitation is particularly relevant for superconducting hydrides, where lattice anharmonicity and NQEs not only dictate the correct geometry but also strongly affect phonon properties, thereby influencing the superconducting $T_\text{c}$. This limitation is also significant for the computation of ice phase transitions at high pressures. \newline
\indent Several methods have been developed to address these challenges. Path-integral molecular dynamics (PIMD) \cite{Ceperley_pimd_1995} and quantum Monte Carlo (QMC) \cite{Mouhat_QMC_PI_2017} approaches offer a rigorous quantum treatment of nuclei but are often prohibitively expensive. The stochastic self-consistent harmonic approximation (SSCHA) method \cite{Errea_PRB_sscha_2014, Monacelli_JPCM_2021_SSCHA}, while often considered the gold standard, is also too costly for large systems. Machine-learned force fields or interatomic potentials reduce computational demands \cite{Lucrezi_MLIP_2023, Ranalli_force_fields_sscha_2023, Kang_mlip_md_2025,Belli_PdCuH2_MLP_2025}, but they often lack the transferability required for reliable predictions across wide composition and pressure ranges and have associated high training costs.
\\
\indent
Herein, we present periodic nuclear–electronic orbital density functional theory (NEO-DFT) \cite{Webb_NEO_2002,Hammes-Schiffer_NEO_Foundations_2021,Xu_periodic_neodft_2022,xu_first-principles_2023,xu_lagrangian_2024} as a computationally efficient and straightforward method for directly incorporating lattice anharmonicity and NQEs into high-pressure system structural optimizations. In the NEO framework, specified nuclei, typically protons or deuterons, are treated quantum mechanically at the same level as the electrons using either DFT or wavefunction methods. For a system with quantum protons, the simplest NEO wavefunction is the product of an electronic determinant, composed of electronic orbitals, and a protonic determinant, composed of protonic orbitals. These orbitals can be obtained by solving the coupled electronic and protonic Hartree--Fock--Roothaan equations self-consistently. Building upon this NEO Hartree--Fock reference, wavefunction methods such as NEO perturbation theory \cite{Fajen_mc-mp4_2021,Hasecke_lmpt_2024}, coupled-cluster theory \cite{Pavosevic-CCSD_NEO_2019,Goudy-triple_exc_2025}, and multireference wavefunction methods \cite{Webb_NEO_2002,Fajen-MulticomponentCASSCF_2021,Malbon-MRCI_2025} have been developed to include correlation effects. 

For calculations of extended materials, NEO-DFT \cite{Pak-NEODFT_electroncorr_2007,Chakraborty-EPfunctionals_2008}, a type of multicomponent DFT \cite{Capitani_nbo_dft_1982,Kreibich_mcdft_2001,Chakraborty_neo_dft_2009}, is a more practical option. Analogous to conventional electronic DFT, NEO-DFT requires the self-consistent solution of coupled electronic and protonic Kohn-Sham equations. In addition to conventional electronic exchange-correlation functionals, accurate NEO-DFT calculations require an electron--proton correlation functional  \cite{yang_development_2017,brorsen_multicomponent_2017,tao_multicomponent_2019}. The formal scaling for NEO-DFT calculations remains the same as that for conventional electronic DFT calculations, rendering it computationally tractable even for relatively large, periodic systems.

By treating specified nuclei quantum mechanically on the same footing as the electrons during the self-consistent field (SCF) procedure, periodic NEO-DFT incorporates NQEs without the need for costly and complicated post-hoc corrections. A straightforward structural optimization with periodic NEO-DFT inherently includes nuclear delocalization, anharmonic zero-point energy, and tunneling effects. Thus, this approach simplifies calculations of crystal structures and phase transitions of hydrogen-rich materials at high pressures. It also enables the efficient search for materials exhibiting specified geometrical properties at high pressures. 

In this work, we demonstrate that periodic NEO-DFT accurately predicts the hydrogen-bond symmetrization pressure in representative solid hydrides, H$_3$S and LaH$_{10}$, while achieving accuracy comparable to a SSCHA calculation at a substantially reduced computational cost. Since the symmetric phase is thought to be dominant within the pressure range exhibiting high T$_\text{c}$ superconductivity, calculating this symmetrization pressure is critical for designing effective high-temperature superconducting hydrides. We then demonstrate that periodic NEO-DFT can describe phase transitions in other high-pressure systems, such as the phase transition from ice VIII to ice X, which also entails hydrogen-bond symmetrization. Periodic NEO-DFT structural optimizations predict the symmetrization pressure for this phase transition of ice and its deuterated counterpart in agreement with experimental measurements at a substantially lower computational cost than other available methods with a similar level of accuracy. These examples highlight the promise of periodic NEO-DFT for understanding and predicting high-pressure phase transitions in a wide range of materials.

\section*{Results and Discussion}
\begin{figure*}[ht]
    \centering
    \includegraphics{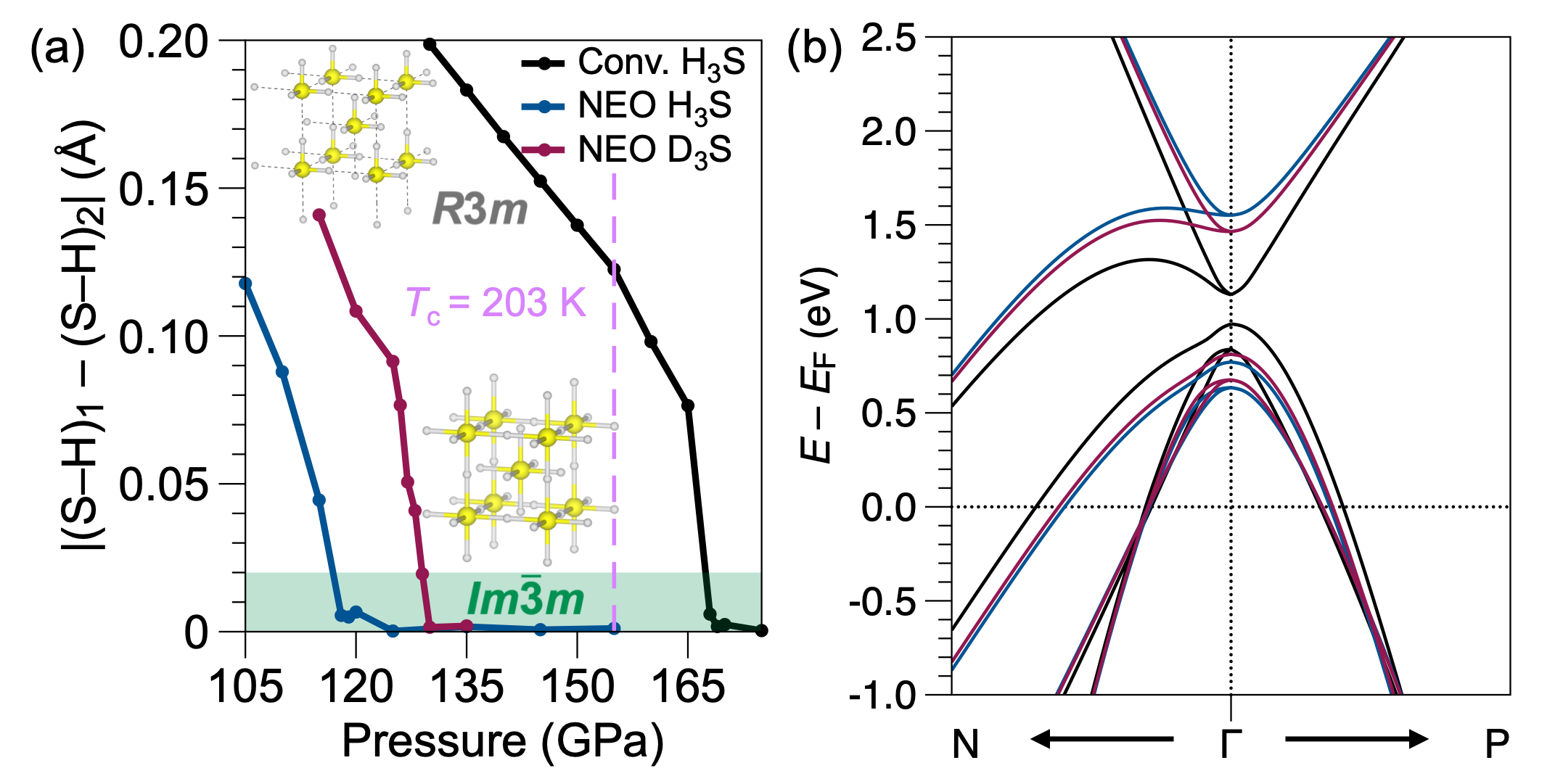}
    \caption{(a) Asymmetric to symmetric phase transition calculated for H$_3$S with conventional periodic DFT (black) and for H$_3$S/D$_3$S (blue/maroon) with periodic NEO-DFT. The green region indicates the region corresponding to the symmetric  {\it{Im}}$\bar{3}${\it{m}} phase. The pink dashed line indicates the pressure corresponding to the highest experimental T$_c$, which occurs at 155 GPa. The $y$-axis corresponds to the difference between S--H distances for two sulfur atoms and the intervening hydrogen. The symmetric phase is defined as this distance being less than 0.02 Angstroms.  Crystal structures in the conventional bcc cell of the {\it{Im}}$\bar{3}${\it{m}} (bottom) and {\it{R}}3{\it{m}} (top) phases of H$_3$S are shown. Solid lines indicate covalent bonds, while the dashed lines in the {\it{R}}3{\it{m}} phase indicate hydrogen bonds. Note that conventional DFT geometry optimizations do not distinguish between H$_3$S and D$_3$S. (b) Electronic band structure of {\it{Im}}$\bar{3}${\it{m}} H$_3$S at 200 GPa performed using conventional DFT and NEO-DFT for H$_3$S and D$_3$S using the same color scheme as part a.}
    \label{fig:H3S_symmetrization}
\end{figure*}

\indent We applied periodic NEO-DFT to two superconducting hydrides, H$_3$S and LaH$_{10}$, and compare the computed hydrogen-bond symmetrization pressure to the values obtained with the state-of-the-art SSCHA method. To move beyond superconducting hydrides, we then used periodic NEO-DFT to compute the phase transition from ice VIII to ice X. In this case, we can compare the computed hydrogen-bond symmetrization pressure to several other computational methods as well as experimental measurements. A key advantage of periodic NEO-DFT over most other methods used for these types of applications is that a single structural optimization inherently includes the NQEs without any additional corrections or processing. Thus, at each pressure studied, only a single periodic NEO-DFT structural optimization is performed for each system. This strategy is not only computationally efficient but is also simple to execute. Moreover, in contrast to conventional periodic DFT geometry optimizations, periodic NEO-DFT geometry optimizations can distinguish between H and D and therefore describe geometric isotope effects.
\subsection*{Superconducting hydrides}
We first examined H$_3$S, a superconducting hydride with a measured high $T_\text{c}$ of 203 K at 155 GPa. In H$_3$S, hydrogen-bond symmetrization marks the transition from the asymmetric {\it{R}}3{\it{m}} phase to the symmetric {\it{Im}}$\bar{3}${\it{m}} phase, which is thought to dominate the pressure range for high-$T_\text{c}$ superconductivity \cite{Mauri_Nature_2016_SH3}. We refer to the pressure at which this process occurs as the symmetrization pressure. Conventional periodic DFT significantly overestimates the symmetrization pressure relative to experimental observations. Our calculations yield a symmetrization pressure of 168 GPa (Fig. \ref{fig:H3S_symmetrization}a), which closely matches the previously reported DFT value of 175 GPa in Ref. \cite{Mauri_Nature_2016_SH3}. In contrast, periodic NEO-DFT calculations predict a lower symmetrization pressure: 118 GPa for H$_3$S and 130 GPa for D$_3$S (Fig. \ref{fig:H3S_symmetrization}a).
These findings are qualitatively consistent with previous SSCHA calculations \cite{Mauri_Nature_2016_SH3}, which predict symmetrization pressures of 103 GPa for H$_3$S and 115 GPa for D$_3$S. Both the periodic NEO-DFT and the SSCHA calculations employed the PBE exchange-correlation functional, which is important for this comparison because the symmetrization pressure is affected by the choice of electronic functional \cite{Mauri_Nature_2016_SH3}.

Although periodic NEO-DFT predicts a slightly higher symmetrization pressure than the SSCHA, both methods identify the symmetric {\it{Im}}$\bar{3}${\it{m}} phase at the experimental pressure of 155 GPa corresponding to the high $T_\text{c}$. We also calculated the electronic band structure for H$_3$S and D$_3$S with periodic NEO-DFT (Fig. \ref{fig:H3S_symmetrization}b) and find qualitatively similar zero-point renormalization effects on the band structure as found previously with other methods that apply complex and expensive perturbative approaches \cite{Sano_PRB_SH3_2016}. In addition, the projected band structures and projected density of states differ in hydrogen character for the bands and density of states for periodic NEO-DFT compared to conventional periodic DFT (see Fig. S1 in Appendix). \newline
\indent To enable a comparison of timings, we repeated the SSCHA calculations using the implementation in the \textsc{SSCHA} Python package \cite{Errea_PRB_sscha_2014, Monacelli_JPCM_2021_SSCHA}. Our results are consistent with those in Ref. \cite{Mauri_Nature_2016_SH3}, with minor differences due to slightly different computational parameters. We emphasize that only the primitive unit cell, consisting of one S and three H, is used for periodic NEO-DFT, whereas SSCHA calculations require a supercell containing 24 S and 72 H. We find a more than 100-fold decrease in computational cost (measured in CPU hours) for the structural optimization of H$_3$S when using periodic NEO-DFT compared to the SSCHA method. \newline
\indent Interestingly, we find that performing the NEO-DFT calculations without the epc17-2 electron--proton correlation functional \cite{brorsen_multicomponent_2017,yang_development_2017} leads to a decrease in the H$_3$S symmetrization pressure by 10 GPa and the D$_3$S symmetrization pressure by 3 GPa, bringing them closer to the SSCHA results (Fig. S3). The electronic band structure also changes when electron--proton correlation is neglected (Fig. S3). A current limitation of NEO-DFT is the neglect of temperature effects, although efforts to include them are ongoing. To enable a consistent comparison, the SSCHA calculations, as well as the NEO-DFT calculations, were performed at 0 K in this work.\newline

\indent LaH$_{10}$, which exhibits a similar nuclear quantum symmetrization effect as H$_3$S, serves as a larger and more complex system to test the capabilities of periodic NEO-DFT. Experiments on lanthanum polyhydrides have reported a weakly pressure-dependent high $T_\text{c}$ of 250 K between 137 and 218 GPa \cite{Geballe_lanthanum_superhydrides_2017,Drodzov_Nature_2019_LaH}, commonly attributed to the symmetric {\it{Fm}}$\bar{3}${\it{m}} phase of LaH$_{10}$ \cite{Mauri_Nature_2020_LaH}. Previous theoretical investigations have revealed a complex energy landscape for this system: conventional DFT identifies multiple minima on the potential energy surface (PES) corresponding to different structural phases \cite{Mauri_Nature_2020_LaH}, whereas inclusion of anharmonic zero-point energy via the SSCHA yields only one minimum, corresponding to the symmetric {\it{Fm}}$\bar{3}${\it{m}} phase. In other words, all other structures relaxed into the symmetric {\it{Fm}}$\bar{3}${\it{m}} phase during the SSCHA structural optimization, consistent with the experimental attribution to this phase \cite{Geballe_lanthanum_superhydrides_2017,Drodzov_Nature_2019_LaH}.\newline   
\indent To illustrate that periodic NEO-DFT captures this behavior, we examined the asymmetric {\it{R}}$\bar{3}${\it{m}} phase and symmetric {\it{Fm}}$\bar{3}${\it{m}} phase of LaH$_{10}$ as a function of applied pressure. These phases can be differentiated by the lattice angle of the primitive unit cell. The symmetrization pressure corresponds to the transition from the asymmetric {\it{R}}$\bar{3}${\it{m}} phase to the symmetric {\it{Fm}}$\bar{3}${\it{m}} phase.

\begin{figure}
    \centering
    \includegraphics{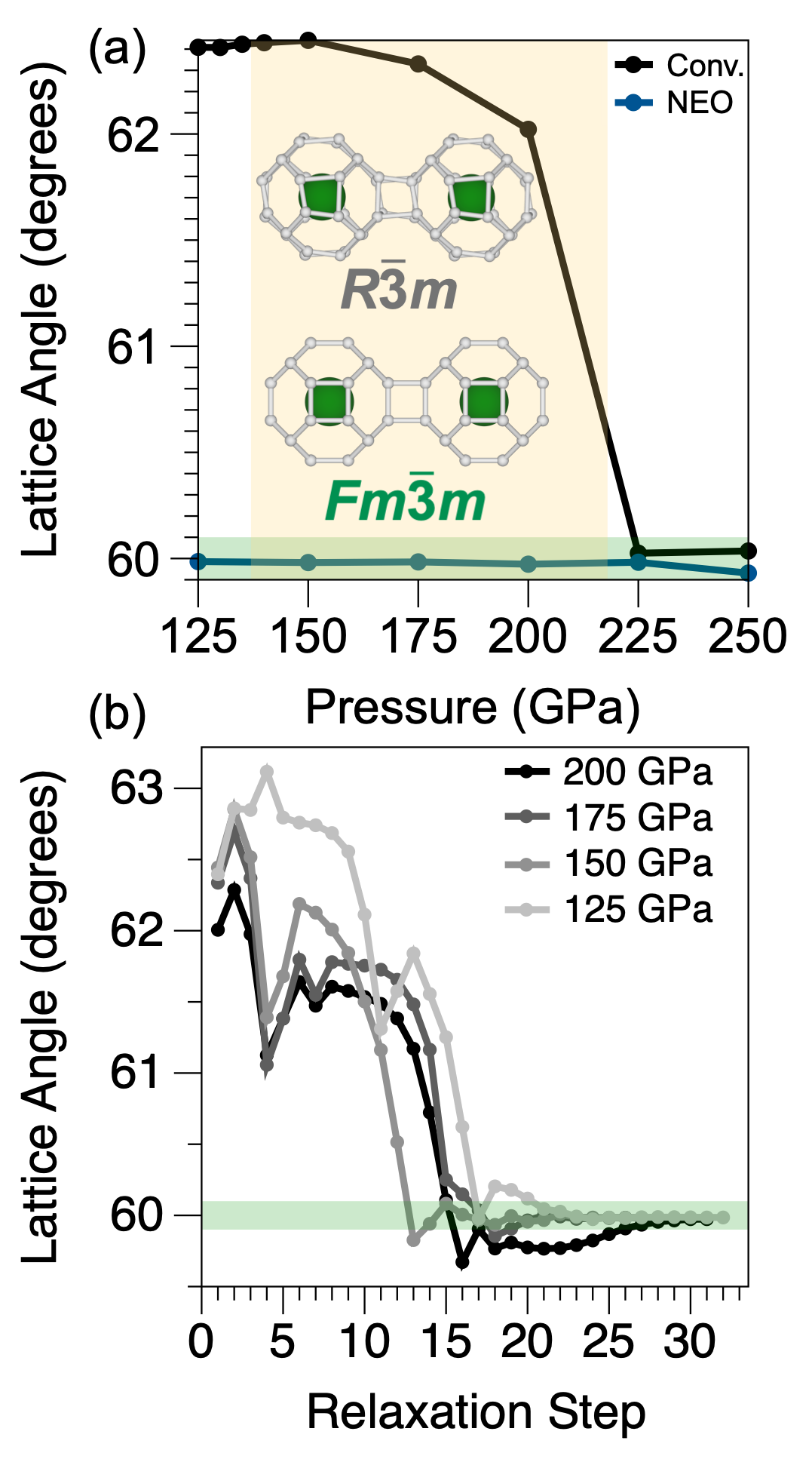}
    \caption{(a) Lattice angle for the lowest energy LaH$_{10}$ structure optimized at different pressures using conventional DFT (black) and NEO-DFT (blue). The asymmetric phase has lattice angles between 62 and 63 degrees, whereas the symmetric phase has lattice angles near 60 degrees. Cage-like structures of LaH$_{10}$ for the {\it{Fm}}$\bar{3}${\it{m}} (symmetric) and {\it{R}}$\bar{3}${\it{m}} (asymmetric) phases are shown. The experimental pressure range where superconductivity is observed, implicating the  symmetric {\it{Fm}}$\bar{3}${\it{m}} structure, is indicated by yellow shading. (b) Lattice angle of LaH$_{10}$ structures at each step of the periodic NEO-DFT structural optimization starting at the asymmetric structure optimized with conventional DFT at different pressures. In both panels, the green shaded area indicates the region corresponding to the symmetric {\it{Fm}}$\bar{3}${\it{m}} structure.}
    \label{fig:LaH10}
\end{figure}
Conventional DFT predicts the symmetrization pressure to be 225 GPa (Fig. \ref{fig:LaH10}a). We performed periodic NEO-DFT optimizations starting at the conventional DFT optimized structures, which corresponded to the asymmetric {\it{R}}$\bar{3}${\it{m}} phase, at pressures of 125, 150, 175, and 200 GPa. At all four pressures, the asymmetric {\it{R}}$\bar{3}${\it{m}} phase evolves to the symmetric {\it{Fm}}$\bar{3}${\it{m}} phase (Fig. \ref{fig:LaH10}b). These results are consistent with the SSCHA results previously,  \cite{Mauri_Nature_2020_LaH} providing further validation for periodic NEO-DFT.
If the periodic NEO-DFT calculation is started from the conventionally optimized {\it{Fm}}$\bar{3}${\it{m}} phase, which is a local minimum on the conventional PES, the structural optimization converges in around half as many steps and remains in the {\it{Fm}}$\bar{3}${\it{m}} phase. 
As shown in the Appendix, the computational time for the periodic NEO-DFT optimization of LaH$_{10}$ is significantly lower than the computational time for the SSCHA optimization of H$_3$S, which is a much smaller system. Presumably the computational advantage is even greater compared to the SSCHA optimization of LaH$_{10}$, although the computational expense for this larger system prevents a direct comparison. \newline 
\subsection*{Ice phase transition}
\begin{figure}
    \centering
    \includegraphics{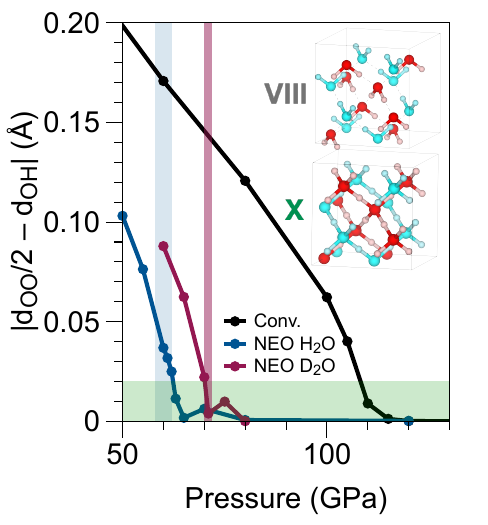}
    \caption{Ice VIII (grey) to ice X (green) phase transition calculated with conventional periodic DFT (black) and NEO-DFT (blue for H, maroon for D) as a function of pressure. In both ice structures, the simulation cell contains 16 water molecules divided into two networks (indicated in red and blue). The green shaded area indicates the region considered to belong to the ice X phase. The blue and maroon shaded areas indicate the experimentally predicted ice VIII to ice X phase transition pressure range for H and D, respectively.}
    \label{fig:ice_neo}
\end{figure}

\indent Ice exhibits a wide variety of both theoretically predicted and experimentally confirmed structures, with one of the richest phase diagrams in nature \cite{engel-2018,loerting-2020}.
At ambient pressure, each hydrogen atom forms a covalent bond with one oxygen atom and a hydrogen bond with another. At higher pressures, the VIII and X phases contain two separate but interwoven hydrogen-bonding networks (Fig. \ref{fig:ice_neo}).
Ice VIII has a preferential direction along which water molecules align, generating a macroscopic net polarization with opposite sign in the two sublattices \cite{ice_vii_viii_1966,Komatsu_transition_7to8,Yoshimura_iceviii,Pruzan_7_to_8}. Ice X, which occurs at higher pressures, exhibits a body-centered cubic (bcc) structure belonging to the space group $Pn\bar{3}m$ \cite{HIRSCH1984142,Bernasconi_1998}.
 \\
\indent In this work, we investigate the ice VIII to ice X phase transition.  Experiments estimate the phase transition pressure for the ice VIII to ice X transition to be between 58 and 62 GPa for H$_2$O \cite{Aoki_exp,Goncharov_exp,Goncharov_exp_2,komatsu-2024} and between 70 and 72 for D$_2$O \cite{Pruzan-1997,Goncharov_exp}. Moreover, this phase transition pressure has been found to be virtually temperature independent in the range 0 to 300 K \cite{Goncharov_exp_2}. From a fundamental perspective, this phase transition is governed by the evolution of the PES experienced by each hydrogen atom situated between two oxygen atoms \cite{lin-2011}.
At lower pressures, the double-well shape of the PES confines hydrogen atoms closer to one oxygen neighbor. At $\sim$60 GPa, however, the barrier drops to a few hundred meV, allowing quantum tunneling and enabling the formation of ice X. Above 100 GPa, the PES becomes a single well, localizing hydrogen atoms exactly at the midpoint, yielding the fully symmetric ice X structure \cite{gsc3-1bz6}. Note that we focus on the pressure associated with the structural change from ice VIII to ice X and do not address the nature of this phase transition or crossover, which has been discussed extensively in the literature \cite{Ackland_icephases_2025}. \\
\indent The phase transition from ice VIII to ice X has been studied with a wide range of theoretical methods. Conventional periodic DFT calculations with a classical treatment of the nuclei predict a transition pressure much higher than the experimental range of 58 to 62 GPa  \cite{Aoki_exp,Goncharov_exp,Goncharov_exp_2,komatsu-2024}. Clearly a quantum treatment of the hydrogen nuclei is necessary to correctly describe this transition.
Previous studies have lowered this transition pressure by including NQEs using a variety of methods, such as the SSCHA \cite{QuantumICE}, PIMD \cite{kang-2013,Kuwahata_PIMD,QuantumICE}, and others \cite{Langevin_vii_to_x,gsc3-1bz6}, but discrepancies between the results obtained with the different methods and experiments remain.
Due to the high computational cost of these calculations, recent studies have utilized machine learning to generate the PES and thereby speed up the calculations \cite{QuantumICE,gsc3-1bz6}.\\ 
\indent Here we show that periodic NEO-DFT structural optimizations accurately predict the transition pressure leading to hydrogen-bond symmetrization for ice VIII to ice X. Each conventional DFT or NEO-DFT calculation was performed on a cell containing 16 water molecules, 8 for each of the two networks, distinguished by color in Figure \ref{fig:ice_neo}, following the prescription from Ref. \cite{QuantumICE}.
The ice VIII and ice X phases are distinguished by the order parameter, $\Delta=|d_{\text{OO}}/2-d_{\text{OH}}|$, for each pair of oxygen atoms and the intervening hydrogen atom in Figure \ref{fig:ice_neo}.
Conventional periodic DFT predicts a transition pressure of 110 GPa. However, the quantum treatment of the hydrogen nuclei using periodic NEO-DFT with the PBE functional leads to a transition pressure of 62 GPa, which agrees with the current experimental estimates \cite{Aoki_exp,Goncharov_exp,Goncharov_exp_2,komatsu-2024}. A previous study using the SSCHA and the PBE functional reported a transition pressure of 51 GPa \cite{QuantumICE}. Similar to H$_3$S, the SSCHA transition pressure agrees better with the NEO-DFT prediction of 50 GPa when no electron--proton correlation functional is used. This finding suggests that differences in electron--proton correlation treatment may be responsible for the minor discrepancy between these two methods.  The results for D$_2$O with NEO-DFT are also shown in Fig. \ref{fig:ice_neo}, and the transition pressure of 71 GPa is in excellent agreement with the experimental value of 72 GPa \cite{Pruzan-1997}. 

The periodic NEO-DFT symmetrization pressures for H/D are summarized and compared to those obtained with other methods that include NQEs in Table \ref{tab:ice_table}.  To enable a direct comparison, all of these calculations used the PBE functional.
Because the choice of electronic functional influences the calculated symmetrization pressure to some extent, we also performed the calculations for H$_2$O with the revPBE-D3 functional \cite{zhang_comment_1998,grimme_semiempirical_2006,grimme_consistent_2010} (Fig. S5), keeping all other computational parameters the same. The calculated transition pressure is similar with this functional.
\begin{table}\centering
\caption{Ice VIII to ice X phase transition pressure (GPa) calculated with different nuclear treatments and the PBE functional.}
\begin{tabular}{lrrrr}
Nuclear & H/D & Transition Pressure & Ref.  \\
\midrule
Experiment & H & 58-62 & \cite{Aoki_exp,Goncharov_exp,Goncharov_exp_2,komatsu-2024} \\
SSCHA & H & 51 & \cite{QuantumICE} \\
PIMD & H & 43 & \cite{QuantumICE} \\
NEO/epc17-2 & H & 62 & this work\\
Experiment & D & 70-72 & \cite{Pruzan-1997,Goncharov_exp} \\
SSCHA & D & 61 & \cite{QuantumICE} \\
NEO/epc17-2 & D & 71 & this work\\
\bottomrule
\end{tabular}
\label{tab:ice_table}
\end{table}
\subsection*{Computational timing comparison}
Although a direct comparison of computational costs between the NEO and SSCHA calculations is not straightforward, two primary factors can lead to significant computational savings when using periodic NEO-DFT for structural optimization compared to the SSCHA. First, the SSCHA requires the use of supercells, implying more expensive SCF calculations. For instance, our SSCHA calculations for H$_3$S employed a 2$\times$2$\times$2 supercell of a three H$_3$S unit cell, leading to 24 times as many H$_3$S stoichiometric units as in the periodic NEO-DFT calculation. Typically, the SSCHA is performed on 2$\times$2$\times$2 or 3$\times$3$\times$3 supercells. Second, SSCHA calculations require generating ensembles of nuclear configurations (often 50 or more), each of which requires a separate SCF calculation. For example, our SSCHA calculations for H$_3$S involved 16 ensembles, each with 50 configurations, resulting in a total of 800 SCF calculations. Some applications of the SSCHA require SCF calculations for tens of thousands of configurations \cite{Taureau_30000configs_2024}. In contrast, a periodic NEO-DFT calculation generally involves a single structural optimization using a primitive unit cell and typically requires on the order of 25 SCF calculations, depending on the initial geometry. A detailed comparison of the NEO and the SSCHA methods is provided in the Methods section, and representative timing data for both approaches are included in Table S1 of the Appendix. Additional techniques to optimize the NEO implementation in FHI-aims \cite{Blum_fhiaims_2009}, such as density fitting \cite{Mejia-Rodriguez_mcdf_2019,Pavosevic_cc_df_2021,hasecke_densityfitting_2023}, Cholesky decomposition \cite{liu_cholesky_2023}, and the nuclear Hartree product representation \cite{Auer_hartreeprod_2010,chow_hartree_2026}, are expected to provide further cost savings for NEO calculations.

\section*{Conclusions}
\indent This work highlights the capability of the periodic NEO-DFT method to accurately incorporate NQEs in high-pressure solid-state materials. The symmetrization pressure predicted by periodic NEO-DFT agrees with experimental measurements and previous theoretical results for the superconducting hydrides, H$_3$S and LaH$_{10}$, as well as the phase transition from ice VIII to ice X. The effects of deuteration on the symmetrization pressures are also described accurately. Furthermore, analysis of the results for ice indicates that electron--proton correlation, which can be included directly in NEO-DFT, improves the agreement of the computed symmetrization pressures with experimental measurements for H$_2$O. 

The power of NEO-DFT is that NQEs, such as nuclear delocalization and anharmonic zero-point energy, are included directly in the  structural optimization within the DFT calculation, removing the need for additional steps and external codes. 
Moreover, periodic NEO-DFT provides significant computational savings compared to the state-of-the-art SSCHA method, corresponding to a factor of at least 100 for the relatively small H$_3$S system and presumably even greater savings for larger systems. The computational savings compared to other state-of-the-art methods, such as PIMD and QMC, are expected to be even more substantial. The efficiency of periodic NEO-DFT will allow for the use of hybrid electronic functionals for systems that would benefit from them and will enable the straightforward study of larger, more complex systems. Thus, this work provides the foundation for future studies of structural phase transitions for a wide range of other solid-state materials.

\section*{Methods}
\subsection*{NEO-DFT}
The NEO approach treats specified nuclei, which are either protons or deuterons in this work, quantum mechanically on the same level as the electrons. Within the NEO-DFT formalism, the coupled electronic and  protonic Kohn–Sham (KS)
equations are expressed as:
\begin{align}   
&\hat{H}_{\mathbf{k}}
^\text{e} \psi_{i,\mathbf{k}}^\text{e}(\mathbf{r}^\text{e})=\left[-\frac{1}{2}\nabla^2_\text{e} + v_{\text{eff}}^\text{e}(\mathbf{r}^\text{e}) \right] \psi_{i,\mathbf{k}}^\text{e}(\mathbf{r}^\text{e})= \epsilon_{i,\mathbf{k}}^\text{e}\psi_{i,\mathbf{k}}^\text{e}(\mathbf{r}^\text{e})\\
& \hat{H}^\text{p} \psi_{i}^\text{p}(\mathbf{r}^\text{p})=\left[-\frac{1}{2m_\text{p}}\nabla^2_\text{p} + v_{\text{eff}}^\text{p}(\mathbf{r}^\text{p}) \right] \psi_{i}^\text{p}(\mathbf{r}^\text{p})= \epsilon_{i}^\text{p}\psi_{i}^\text{p}(\mathbf{r}^\text{p})
\end{align}
where $m_\text{p}$ is the proton mass, $\psi_{i,\mathbf{k}}^\text{e}$ and $\epsilon_{i,\mathbf{k}}^\text{e}$ are the KS orbitals and
eigenvalues, respectively, for the electrons, and $\psi_j^\text{p}$ and $\epsilon_i^\text{p}$ are the KS
orbitals and eigenvalues, respectively, for the quantum protons. The effective
potentials are defined as:
\begin{align}
&
\begin{split}
v_{\text{eff}}^\text{e}(\mathbf{r}^\text{e})=&v_{\text{ext}}(\mathbf{r}^\text{e}) + v_{\text{es}}^\text{e}(\mathbf{r}^\text{e}) - v_{\text{es}}^\text{p}(\mathbf{r}^\text{e})+\\
&+ \frac{\delta E_{\text{xc}}^\text{e}[\rho^\text{e}]}{\delta\rho^\text{e}}+\frac{\delta E_{\text{epc}}[\rho^\text{e},\rho^\text{p}]}{\delta\rho^\text{e}}
\end{split}\\
& \begin{split}
v_{\text{eff}}^\text{p}(\mathbf{r}^\text{p})=&-v_{\text{ext}}(\mathbf{r}^\text{p}) - v_{\text{es}}^\text{e}(\mathbf{r}^\text{p}) + v_{\text{es}}^\text{p}(\mathbf{r}^\text{p})+ \\ 
&+ \frac{\delta E_{\text{xc}}^\text{p}[\rho^\text{p}]}{\delta\rho^\text{p}}+\frac{\delta E_{\text{epc}}[\rho^\text{e},\rho^\text{p}]}{\delta\rho^\text{p}}
\end{split}
\end{align}
where $v_{\text{ext}}$, $v_{\text{es}}^\text{e}$, and $v_{\text{es}}^\text{p}$ are electrostatic potentials for the classical nuclei, electrons, and quantum protons, respectively. $E_{\text{xc}}^\text{e}$ and $E_{\text{xc}}^\text{p}$ are the exchange–correlation energies for the electrons and the quantum protons, corresponding to electron-electron and proton-proton interactions, respectively. 
$E_{\text{epc}}$ is the electron--proton correlation energy, which is a functional of both the electronic density and the protonic density. 
The ground state of the multicomponent system of electrons and quantum protons is obtained by solving these equations self-consistently in the presence of the electrostatic potential from the classical nuclei.

\indent The implementation of periodic NEO-DFT in the FHI-aims code \cite{Blum_fhiaims_2009} adopts numerically tabulated atom-centered orbitals (NAOs) for the electronic basis sets and Gaussian-type orbital basis sets \cite{yu_development_2020} for the quantum protons and deuterons. Herein, the ``light" electronic basis set \cite{Blum_fhiaims_2009} and PB4-F2 protonic basis set \cite{yu_development_2020} were used for all NEO calculations. For calculations with deuterium, the exponents of the protonic basis set were scaled by $\sqrt{2}$ to account for the mass difference. The PBE electronic exchange-correlation functional \cite{Perdew_PRL_1996_PBE} was used for both NEO and conventional DFT calculations. The epc17-2 electron--proton correlation functional \cite{brorsen_multicomponent_2017,yang_development_2017} was used in the NEO calculations, and the protonic exchange-correlation functional is simply the exact Hartree--Fock exchange, as proton-proton correlation has been shown to be negligible \cite{pavosevic_rev_2020}. The basis set convergence is shown in Fig. S2, and the impact of the epc17-2 functional is shown in Figs. S3 and S4.

For the structural optimizations, the forces were converged to 0.005 eV Å$^{-1}$. The atomic zero-order regular approximation (ZORA) scalar-relativistic approximation as implemented in FHI-aims was used to capture relativistic effects \cite{Blum_fhiaims_2009}. A 24$\times$24$\times$24 \textit{k}-point mesh was used for H$_3$S, a 32$\times$32$\times$32 \textit{k}-point mesh was used for LaH$_{10}$ , and a 5$\times$5$\times$5 \textit{k}-point mesh was used for ice. The effect of external pressure was included by applying hydrostatic pressure and relaxing the unit cell until the desired pressure was reached \cite{knuth_stress_2015}.  \newline
%
\indent The computational cost of a periodic NEO-DFT structural optimization can be estimated as $N_\text{steps} \times T_{\rm scf}^{\rm NEO}$, where $N_\text{steps}$ is the number of geometry optimization steps necessary to obtain convergence and $T_{\text{scf}}^{\text{NEO}}$ is the computational time of a NEO-DFT SCF calculation. The latter scales on the same order as a conventional electronic DFT calculation, although nuclear basis functions are required because the protons are also treated quantum mechanically. Typically, the additional cost for NEO calculations is relatively low because the electronic basis functions greatly outnumber the protonic basis functions.
\subsection*{SSCHA}
The SSCHA is an SCF method that captures quantum, thermal, and anharmonic nuclear effects within the Born--Oppenheimer approximation. The method relies on the Gibbs--Bogoliubov variational principle to approximate the free energy of the ionic Hamiltonian $\mathcal{F}_H$, which is computed from a trial density matrix $\tilde{\rho}$ \cite{Monacelli_JPCM_2021_SSCHA}:

\smallskip
\begin{equation}
\mathcal{F}_H\left[\rho\right] \leq \mathcal{F}_H\left[\tilde{\rho}\right]
\end{equation}
\smallskip
The trial density matrix is constrained to multidimensional Gaussian distributions $\tilde{\rho}_{\bm{\mathcal{R}},\bm\Phi}$, parameterized by $\bm{\mathcal{R}}=\{\bm{\mathcal{R}_I}\}_{I=1}^N$, a $3N$-dimensional vector representing the average atomic positions, and $\bm\Phi$, a $3N\times3N$ positive-definite matrix that encodes quantum and thermal fluctuations around these positions.
The SSCHA evaluates the free energy from an ensemble of supercell configurations sampled stochastically from the probability distribution defined by the trial density matrix. The minimization starts with initial $\bm{\mathcal{R}}^{(0)}$ and $\bm\Phi^{(0)}$, typically derived from density functional perturbation theory calculations. An external \textit{ab initio} code computes energies, forces, and stress tensors for each configuration. These quantities are used to evaluate $\mathcal{F}_H[\bm{\mathcal{R}},\bm\Phi]$ and its derivatives via Monte Carlo integration, updating $\bm{\mathcal{R}}$ and $\bm{\Phi}$ by preconditioned gradient descent. This process is iterated until the free energy gradients converge, often requiring many ensembles. Upon convergence, in addition to the temperature-dependent $\mathcal{F}$, $\bm{\mathcal{R}}_{\mathrm{eq}}$, and $\bm\Phi_{\mathrm{eq}}$, the algorithm also computes the forces and the stress tensor, enabling structure relaxation at finite temperature and pressure, with full inclusion of quantum and thermal ionic fluctuations. \\
\indent For direct comparison to the NEO-DFT H$_3$S results and to enable a comparison of timings, we employed the SSCHA method implemented in the \textsc{SSCHA} Python package \cite{Errea_PRB_sscha_2014, Monacelli_JPCM_2021_SSCHA} and interfaced with Quantum ESPRESSO \cite{quantumespresso_1,quantumespresso_2}. For these SSCHA calculations, the PBE electron exchange-correlation functional \cite{Perdew_PRL_1996_PBE} was used with norm-conserving Vanderbilt pseudopotentials \cite{Hamann_PRB_2017_ONCV} and a kinetic energy cutoff of 80 Ry. An 8$\times$8$\times$8 \textit{k}-point mesh was used for the unit cell containing 3 H$_3$S units. The \textit{k}-point mesh was scaled accordingly for the 2$\times$2$\times$2 supercell (24 H$_3$S units) used in the ensemble calculations. 50 atomic configurations were generated in each ensemble. The SSCHA calculation was performed at 0 K. 
The computational cost of a SSCHA calculation scales roughly as $N_\text{ens} \times N_\text{ind} \times T_{\rm scf}^{\rm DFT,sc}$, where $N_\text{ens}$ is the number of ensembles (or iterations) required for convergence, $N_\text{ind}$ is the ensemble size (typically 50–100 configurations), and $T_{\rm scf}^{\rm DFT,sc}$ is the computational time of a conventional DFT SCF calculation on the supercell (sc).
\subsection*{Comparison of periodic NEO-DFT and SSCHA calculations} 
Here we summarize a few fundamental differences between periodic NEO-DFT and the SSCHA. The NEO-DFT method does not currently allow for finite-temperature effects in the structural optimization, which is possible with the SSCHA. In contrast to the SSCHA, the NEO-DFT method treats the quantum nuclei and electrons on the same level without any Born--Oppenheimer separation between them. Thus, the nuclear quantum effects are included directly during the NEO-DFT SCF procedure and during the structural optimization rather than requiring a separate procedure. Moreover, the NEO-DFT method employs more flexible protonic basis sets when compared to the SSCHA method, which only utilizes a single Gaussian to represent each protonic orbital. The NEO-DFT method can also directly include electron--proton correlation through an electron--correlation functional. As a result, NEO-DFT allows for systematic investigation into the impact of electron--proton correlation by performing the calculations with and without the  epc17-2 functional. Although the SSCHA may be implicitly capturing some electron--proton correlation through random sampling of nuclear configurations, this contribution cannot be systematically investigated or controlled.

\subsection*{Supporting Information Appendix (SI)}
Representative timings and additional calculations.

\section*{Acknowledgments}
The NEO-DFT work was funded by National Science Foundation Grant No. CHE-2408934 (S.H.-S.). L.E.S recognizes support from a National Science Foundation Graduate Research Fellowship under grant DGE-2444107. This work used Expanse at the San Diego Supercomputer Center through allocation MCB120097 from the Advanced Cyberinfrastructure Coordination Ecosystem: Services \& Support (ACCESS) program, which is supported by U.S. National Science Foundation grants \#2138259, \#2138286, \#2138307, \#2137603, and \#2138296 \cite{boerner_access_2023}. L.B. acknowledges support from Fondo Ateneo Sapienza 2019-22, and funding from the European Union - NextGenerationEU under the Italian Ministry of University and Research (MUR), “Network 4 Energy Sustainable Transition - NEST” project (MIUR project code PE000021, Concession Degree No. 1561 of October 11, 2022) - CUP: B53C22004070006.

\printbibliography

\end{document}


\title{Supporting Information: Capturing nuclear quantum effects in high-pressure superconducting hydrides and ice with nuclear--electronic orbital theory}

\author[a]{Logan E. Smith}
\affil[a]{Department of Chemistry, Princeton University, Princeton, New Jersey 08544, United States}
\author[b,c]{Paolo Settembri}
\affil[b]{PSI Center for Scientific Computing, Theory and Data, Paul Scherrer Institute, 5232 Villigen PSI, Switzerland}
\affil[c]{Department of Physical and Chemical Sciences, University of L'Aquila, Via Vetoio, 67100 L'Aquila, Italy}
\author[d]{Alessio Cucciari}
\author[d]{Lilia Boeri}
\affil[d]{Dipartimento di Fisica, Sapienza - Universit\`a di Roma, 00185 Rome, Italy}
\author[c,e]{Gianni Profeta}
\affil[e]{CNR-SPIN L'Aquila, Via Vetoio, 67100 L'Aquila, Italy}
\author[a]{Sharon Hammes-Schiffer}

\maketitle

\begin{table}\centering
    \begin{threeparttable}
        \caption{Representative timings\tnote{a} for conventional periodic DFT, periodic NEO-DFT, and the SSCHA calculations}
        \begin{tabular}{l l l l}
            Species & Method & Calculation Type\tnote{b} & CPU Hours\\\hline
            H$_3$S & Conventional & SCF & 0.16 \\\hline
            H$_3$S & NEO & SCF & 1.5 \\\hline
            H$_3$S & SSCHA\tnote{c} & SCF & 4.3\\\hline
            H$_3$S & NEO & Optimization & 20 \\\hline
            H$_3$S & SSCHA\tnote{c} & Optimization & 3180 \\\hline
            LaH$_{10}$ & NEO & SCF & 15 \\\hline
            LaH$_{10}$ & NEO & Optimization & 450 \\\hline
            Ice & NEO & SCF & 32 \\\hline
            Ice & NEO & Optimization & 1280 \\\hline
        \end{tabular}
         \label{tab:Timings}
         \begin{tablenotes}
         \item[a] Timings given in CPU hours for calculations performed on AMD EPYC 7742 CPUs with 16 CPUs for the conventional periodic DFT and periodic NEO-DFT calculations and 48 CPUs for the SSCHA calculations. Both conventional periodic DFT and periodic NEO-DFT calculations were performed with FHI-aims, whereas the SSCHA calculations were performed with the SSCHA Python package interfaced with Quantum ESPRESSO. The nuclear Hartree product representation has been shown to significantly decrease the number of iterations required for convergence, often leading to a factor of 5--10 speed-up for NEO-DFT SCF calculations with multiple quantum nuclei \cite{chow_hartree_2026}, but it has not been implemented yet in FHI-aims. Thus, these NEO timings should be viewed as an upper limit that will decrease significantly with this modification. However, a clear and substantial computational savings with the NEO method over the SSCHA for geometry optimizations can be observed.
         \item[b] The structural optimizations started in the asymmetric phase optimized with conventional DFT at 155 GPa for H$_3$S, 200 GPa for LaH$_{10}$, and 60 GPa for ice, and relaxed to the symmetric phase with either periodic NEO-DFT or the SSCHA. The SCF calculations were started from the resulting optimized structure with the initial electron density obtained from the superposition of atomic densities (SAD) method. 
         \item[c] The SSCHA calculations were performed with a supercell containing 24 H$_3$S units, resulting in an increase in computational time for the SCF calculation compared to the periodic NEO-DFT calculations, which were performed with a primitive unit cell. In addition, the need for SCF calculations for each nuclear configuration in the ensembles leads to a significantly increased computational cost for structural optimizations with the SSCHA compared to a single structural optimization with periodic NEO-DFT.
         \end{tablenotes}
    \end{threeparttable}
\end{table}

\newpage

\begin{figure*}[h]
    \centering
    \includegraphics{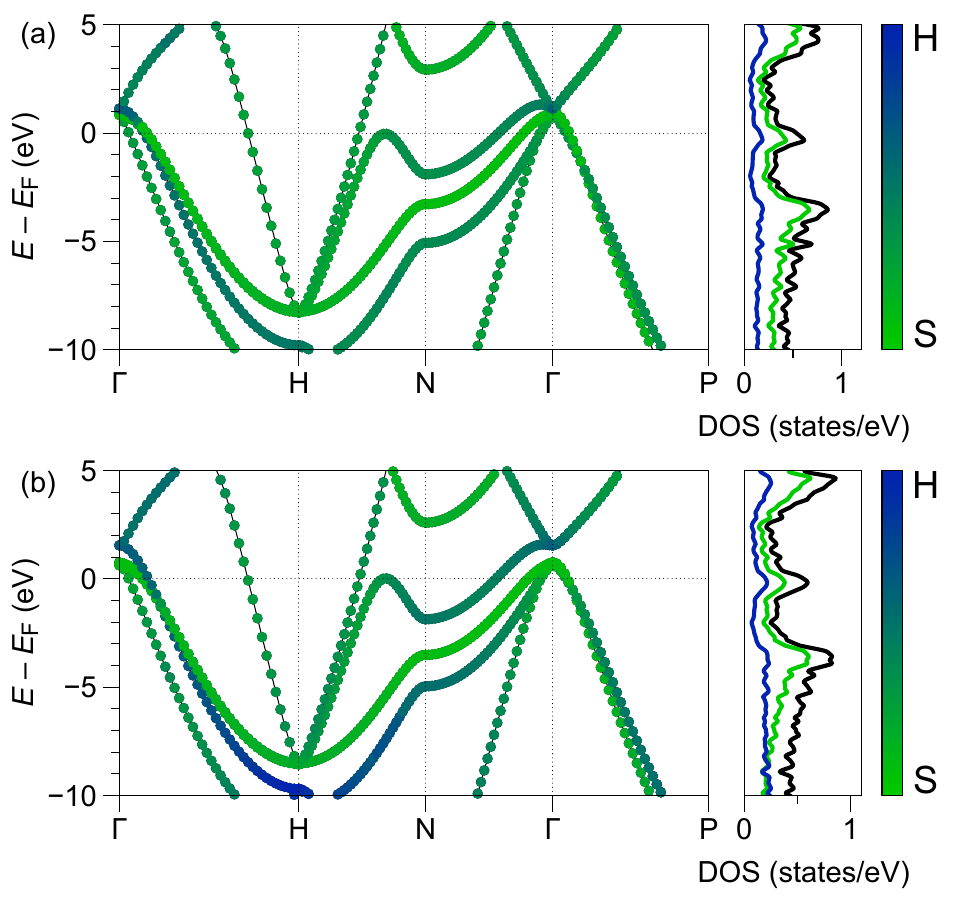}
    \caption{Projected band structures and density of states of the {\it{Im}}$\bar{3}${\it{m}} structure of H$_3$S at 200 GPa for (a) conventional periodic DFT and (b) periodic NEO-DFT. Blue indicates H character and green indicates S character, while the black line indicates the total density of states. The main difference is a clear increase in the hydrogen character of the electronic states along the $\Gamma$-N band close to the Fermi energy, and the bands passing through H from -8 to -10 eV. }
    \label{fig:Bands_char}
\end{figure*}

\newpage

\begin{figure}
    \centering
        \includegraphics[width=\linewidth]{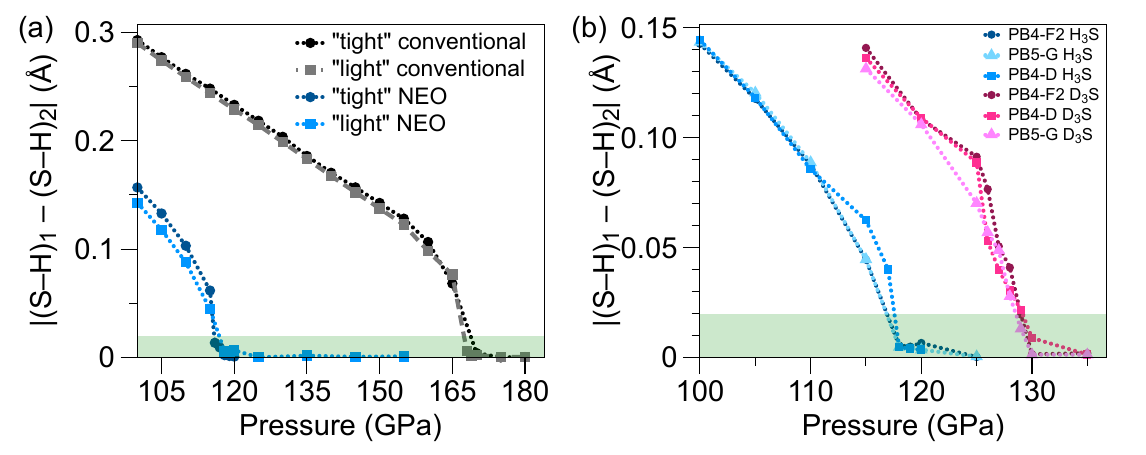}
    \caption{Basis set dependence of asymmetric to symmetric phase transition for H$_3$S and D$_3$S computed with  conventional periodic DFT and periodic NEO-DFT. (a) Comparison using ``light" and ``tight" electronic basis sets for H$_3$S. The PB4-F2 protonic basis set was used. (b) Comparison using three different protonic basis sets. For deuterium, the exponents were multiplied by $\sqrt{2}$ to account for the mass difference. The ``light" electronic basis set was used. In both panels, the PBE electronic functional and epc17-2 electron-proton correlation functional were used.
    The green region indicates the 0.02 Å cutoff for the {\it{Im}}$\bar{3}${\it{m}} phase. The ``light" electronic basis set with PB4-F2 was used for all systems investigated in this work because it retains the accuracy of the larger basis set with significant computational speed-up.}
    \label{fig:basiscompare}
\end{figure}

\newpage

\begin{figure*}[b]
    \centering
    \includegraphics{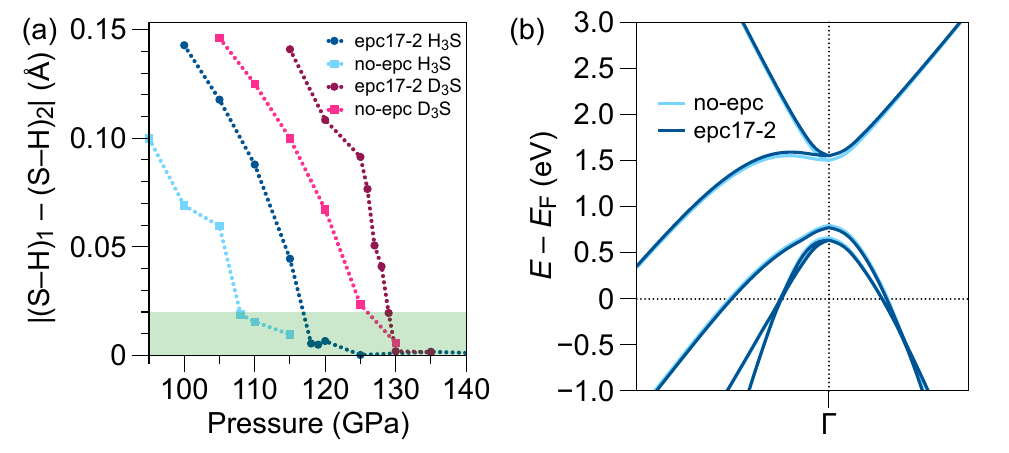}
    \caption{Effect of the electron--proton correlation functional on periodic NEO-DFT calculations of the phase transition and band structure for H$_3$S. (a) Phase  transition from the asymmetric phase to the symmetric phase of H$_3$S (blue) and D$_3$S (maroon) calculated with (dark circles) and without (light squares) the epc17-2 electron--proton correlation functional. The green region indicates the 0.02 Å cutoff for the symmetric {\it{Im}}$\bar{3}${\it{m}} phase. (b) Band structure of {\it{Im}}$\bar{3}${\it{m}} H$_3$S at 200 GPa calculated with (dark blue) and without (light blue) the epc17-2 electron--proton correlation functional. For both panels, the ``light" electronic basis set and PB4-F2 protonic basis set were used. For deuterium, the exponents were multiplied by $\sqrt{2}$ to account for the mass difference.}
    \label{fig:EPC_effects}
\end{figure*}

\newpage

\begin{figure}[h]
    \centering \includegraphics{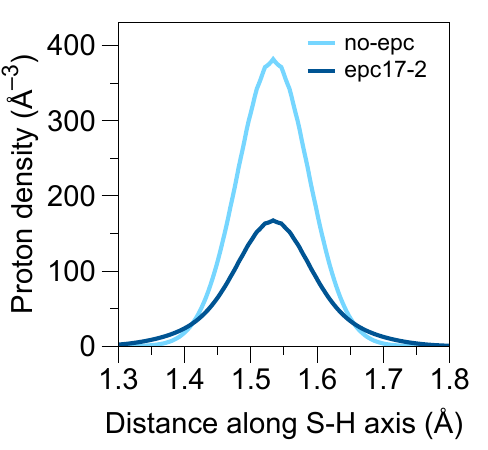}
    \caption{Protonic density along the S-H bond in Im$\bar{3}$m H$_{3}$S at 150 GPa calculated with (dark blue) and without (light blue) the epc17-2 electron-proton correlation functional. The protonic density is much too localized when electron-proton correlation is not included. }
    \label{fig:charge}
\end{figure}

\newpage

\begin{figure}[h]
    \centering
    \includegraphics{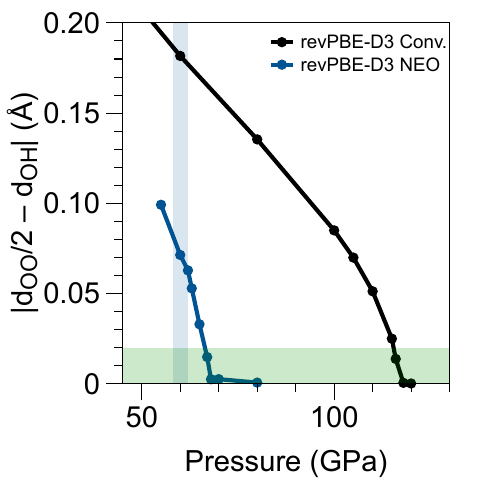}
    \caption{Ice VIII (grey) to ice X (green) phase transition calculated with conventional periodic DFT (black) and NEO-DFT (blue) as a function of pressure using the revPBE-D3 electronic functional. In both ice structures, the simulation cell contains 16 water molecules divided into two networks. The green shaded area indicates the region considered to belong to the ice X phase. These calculations with the revPBE-D3 and epc17-2 functionals produce a transition pressure of $\sim$116 GPa with conventional periodic DFT and $\sim$66 GPa with periodic NEO-DFT. The blue shaded area indicates the experimentally predicted ice VIII to X phase transition pressure range for H$_2$O.}
    \label{fig:revpbe}
\end{figure}

\clearpage
\printbibliography